\documentclass{article}
\usepackage{arxiv}
\usepackage[utf8]{inputenc} 
\usepackage[T1]{fontenc}    
\usepackage{hyperref}       
\usepackage{url}         
\usepackage{booktabs}    
\usepackage{amsfonts}      
\usepackage{nicefrac}      
\usepackage{microtype}     
\usepackage{lipsum}
\usepackage{float}
\usepackage{graphicx}
\usepackage{caption}
\usepackage{subcaption}
\usepackage{titlesec}
\usepackage{multirow}
\usepackage{amsmath}
\usepackage{soul,color}
\usepackage[section]{placeins}

\usepackage[nodisplayskipstretch]{setspace}

\title{Contact map based crystal structure prediction using global optimization}
\author{
  Jianjun Hu*\\
  Department of Computer Science and Engineering\\
  University of South Carolina\\
  Columbia, SC 29201 \\
\texttt{jianjunh@cse.sc.edu} 
    \And
 Wenhui Yang, Rongzhi Dong, Yuxin Li, Xiang Li, Shaobo Li\\
 School of Mechanical Engineering\\
  Guizhou University \\
  Guiyang China 550050 \\
  \And
    Edirisuriya MD Siriwardane\\
  Department of Computer Science and Engineering\\
  University of South Carolina\\
  Columbia, SC 29201 \\
}
\titlespacing {\section}
{0pt}{5.5ex plus 1ex minus .2ex}{3.3ex plus .2ex}
\begin{document}
\maketitle

\begin{abstract}
Crystal structure prediction is now playing an increasingly important role in the discovery of new materials or crystal engineering. Global optimization methods such as genetic algorithms (GA) and particle swarm optimization have been combined with first principle free energy calculations to predict crystal structures given composition or only a chemical system. While these approaches can exploit certain crystal patterns such as symmetry and periodicity in their search process, they usually do not exploit the large amount of implicit rules and constraints of atom configurations embodied in the large number of known crystal structures. They currently can only handle crystal structure prediction of relatively small systems. Inspired by the knowledge-rich protein structure prediction approach, herein we explore whether known geometric constraints such as the atomic contact map of a target crystal material can help predict its structure given its space group information. We propose a global optimization based algorithm, CMCrystal, for crystal structure (atomic coordinates) reconstruction based on atomic contact maps. Based on extensive experiments using six global optimization algorithms, we show that it is viable to reconstruct the crystal structure given the atomic contact map for some crystal materials but more geometric or physicochemical constraints are needed to achieve the successful reconstruction of other materials

\end{abstract}

\section{Introduction}








Accurate computational prediction of crystal materials structures have a variety of important applications. It can be used for computational discovery of novel functional materials has big potential in transforming a variety of industries such as cell phone batteries, electric vehicles, quantum computing hardware, catalysts\cite{oganov2019structure}. Compared to traditional Edisonian or trial-and-error approaches which usually strongly depend on the expertise of the scientists, computational materials discovery has the advantage of efficient search in the vast chemical design space. Togethre with inverse design\cite{zunger2018inverse,kim2020inverse} and generative machine learning models\cite{dan2019generative,bradshaw2019model,kim2020inverse,noh2019inverse,ren2020inverse}, crystal structure prediction (CSP) \cite{glass2006uspex,oganov2011modern,oganov2019structure,kvashnin2019computational} is among the most promising approaches for new materials discovery. Crystal structure prediction also has important applications in crystal engineering, which is concerned with design and synthesis of molecular solid state structures with desired properties. For example, CSP allows us to conduct mutagenesis experiments to examine how composition changes may affect the structural mutations in terms of lattice constant changes or symmetry breaking. Crystal structure prediction can also be used to augment the X-ray diffraction (XRD) based crystal structure determination via space group identification \cite{oviedo2019fast} or providing initial parameters for the XRD based Rietveld refinement method for structure determination \cite{ozaki2020automated}

In a standard crystal structure prediction (CSP) problem\cite{lyakhov2013new}, one has to find a crystal structure with the lowest free energy for a given chemical composition (or a chemical system such as Mg-Mn-O with variable composition) at given pressure–temperature conditions
\cite{oganov2019structure}.  With the crystal structure of a chemical substance, many physicochemical properties can be predicted reliably and routinely using first-principle calculation or machine learning models \cite{xie2018crystal}. It is assumed that lower free energy corresponds to the more stable arrangement of atoms. The CSP approach for new materials discovery is especially appealing due to the efficient sampling algorithm that generates diverse chemically valid candidate compositions with low free energies\cite{dan2019generative}.  CSP algorithms based on evolutionary algorithms \cite{oganov2006crystal} and particle swarm optimization \cite{wang2015materials} have led to a series of new materials discoveries \cite{oganov2011evolutionary,oganov2019structure,wang2020calypso}. However, these global free energy search based algorithms have a major obstacle that limits their successes to relative simple crystals \cite{oganov2019structure,zhang2017materials} (mostly binary materials with less than 20 atoms in the unit cell\cite{oganov2019structure,wang2020calypso}) due to their dependence on the costly DFT calculations of free energies for sampled structures. With limited DFT calculations budget, how to efficiently sample the atom configurations becomes a key issue \cite{oganov2011evolutionary,lyakhov2013new}. To improve the sampling efficiency, a variety of strategies have been proposed such as exploiting symmetry\cite{pretti2020symmetry} and pseudosymmetry\cite{lyakhov2013new}, smart variation operators, clustering, machine-learning interatomic potentials with active learning \cite{podryabinkin2019accelerating}. Several heuristic ideas that exploit known structures in the databases have also been proposed \cite{schon2014can}. However, the scalability of these approaches remains an unsolved issue.

Recently, generative machine learning models have been emerging as a novel approach to generate new materials including generative adversarial networks (GAN) approach for both chemical composition discovery \cite{dan2020generative} and crystal structure generation for a given chemical system \cite{kim2020generative} and autoencoder based models for crystal structure generation\cite{noh2019inverse,ren2020inverse}. Compared to global free energy approaches in CSP, these methods can take advantage of the implicit composition, atomic configuration rules, and constraints embodied in the large number of known crystal structures that can be learned by the deep neural network models. Using neural networks to implicitly learn such rules may lead to more efficient sampling of the search space \cite{dan2020generative}. 

\begin{figure}[h]
	\centering
	\begin{subfigure}{.45\textwidth}
		\includegraphics[width=\textwidth]{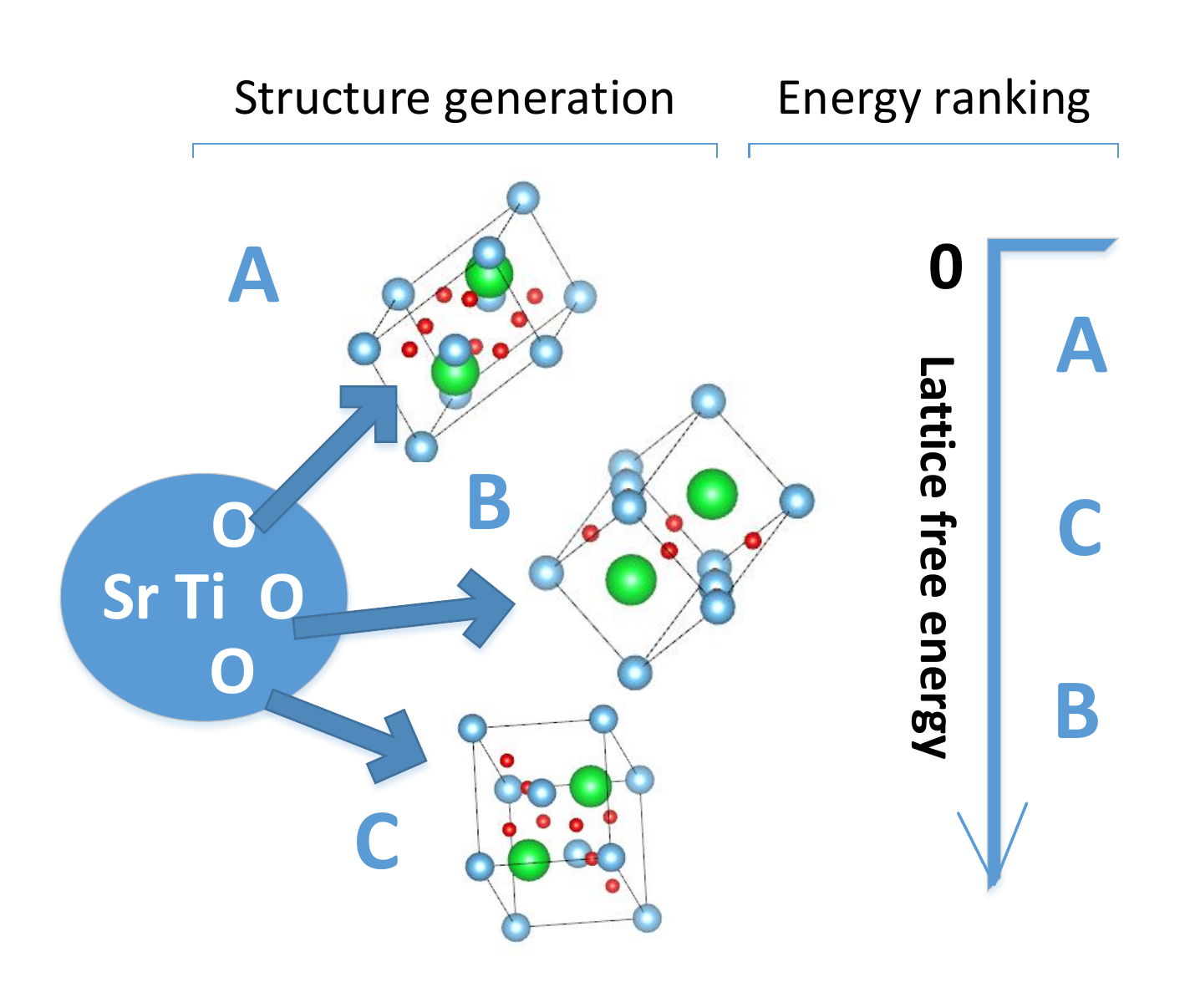}
		\caption{Conventional free energy minimization based CSP}
		\vspace{3pt}
	\end{subfigure}
	\begin{subfigure}{.45\textwidth}
		\includegraphics[width=1.1\textwidth]{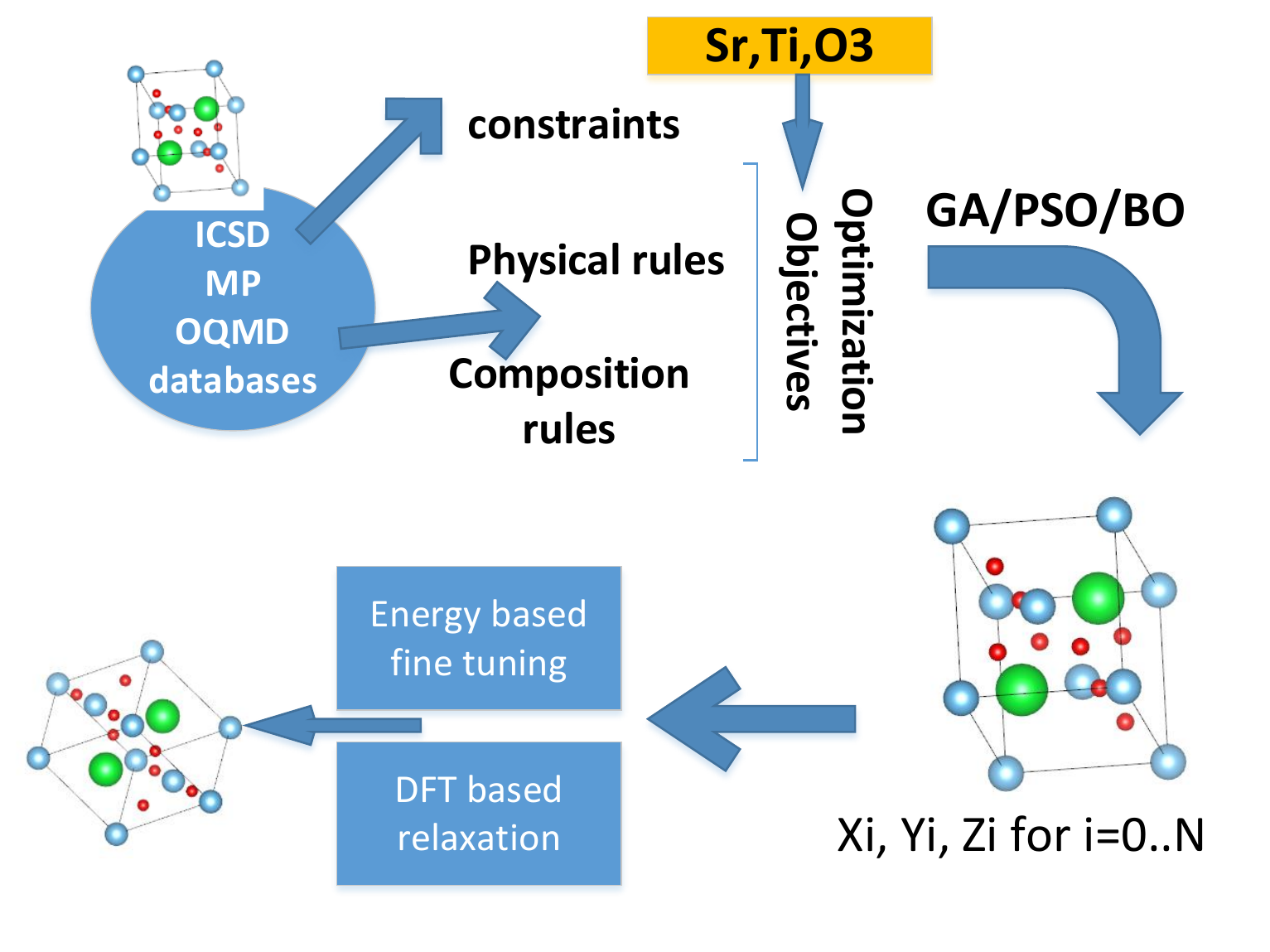}
		\caption{Knowledge-rich CSP}
		\vspace{3pt}
	\end{subfigure}

	\caption{Comparison of traditional crystal structure prediction (CSP) and knowledge-guided CSP. Conventional CSP is limited by its dependence on expensive DFT calculations of free energies while knowledge-rich CSP exploits chemical rules and geometric or physical constraints from known crystals to guide the structure search.}
	\label{fig:comparison}
\end{figure}

Herein we explore a new knowledge-rich approach for crystal structure prediction, which is inspired by the recent success of deep learning approaches for protein structure prediction (PSP)\cite{zheng2019deep} led by the famous AlphaFold \cite{senior2020improved} algorithm from Google DeepMind. In the PSP problem, one has to predict the 3D tertiary structure of a protein given only its amino acid sequence. The latest approach uses deep learning to predict the contact maps\cite{emerson2017protein} or distance matrix\cite{senior2020improved}, which can then be used to reconstruct the full three-dimensional (3D) protein structure with high accuracy\cite{vendruscolo1997recovery}. In this paper, we are exploring how we can use global optimization algorithms to reconstruct the atomic configuration for a given composition based on its space group and the atomic contact map. The idea is that we can exploit the rich atom interaction distribution or other geometric patterns or motifs \cite{zhu2017efficient} existing in the large number of known crystal structures to predict the atomic contact map. The space group of crystal structures can also be predicted using a variety of prediction algorithms \cite{zhao2020machine,liang2020cryspnet} or be inferred from domain knowledge \cite{cabeza2007space}. In \cite{liang2020cryspnet}, the top-3 accuracy for space group prediction ranges from 81\% to 100\% given its Bravais lattice, which can also be predicted using composition features with up to 84\% accuracy. With the predicted contact map and the space group, we investigate whether global optimization algorithms such as GAs and CMA-ES methods can be used for predicting its crystal structure. The comparison of the main differences between conventional CSP and knowledge-rich CSP is shown in Figure\ref{fig:comparison}.

Our contributions can be summarized as follows:

\begin{itemize}
  \item We propose a new approach for crystal structure prediction using the atomic contact map as a knowledge-rich methodology for solving CSP problems.
  \item We define a series of benchmark test cases for testing global optimization algorithms to reconstruct the atomic configurations from atomic contact maps
  \item we conduct extensive evaluations of how different global optimization algorithms perform in contact map based crystal structure prediction.
\end{itemize}

\section{Materials and Methods}

\subsection{Problem formulation: knowledge-rich contact map based CSP }

\begin{figure}[h]
	\centering
	\begin{subfigure}{.45\textwidth}
		\includegraphics[width=\textwidth]{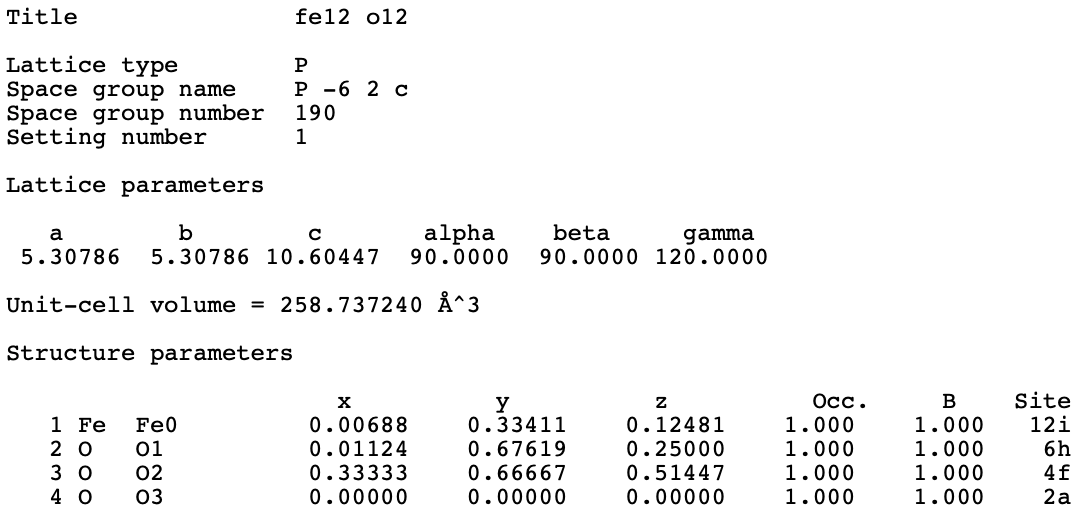}
		\caption{Cif file of crystal material Fe\textsubscript{12}O\textsubscript{12}}
		\vspace{3pt}
	\end{subfigure}
	\begin{subfigure}{.45\textwidth}
		\includegraphics[width=0.6\textwidth]{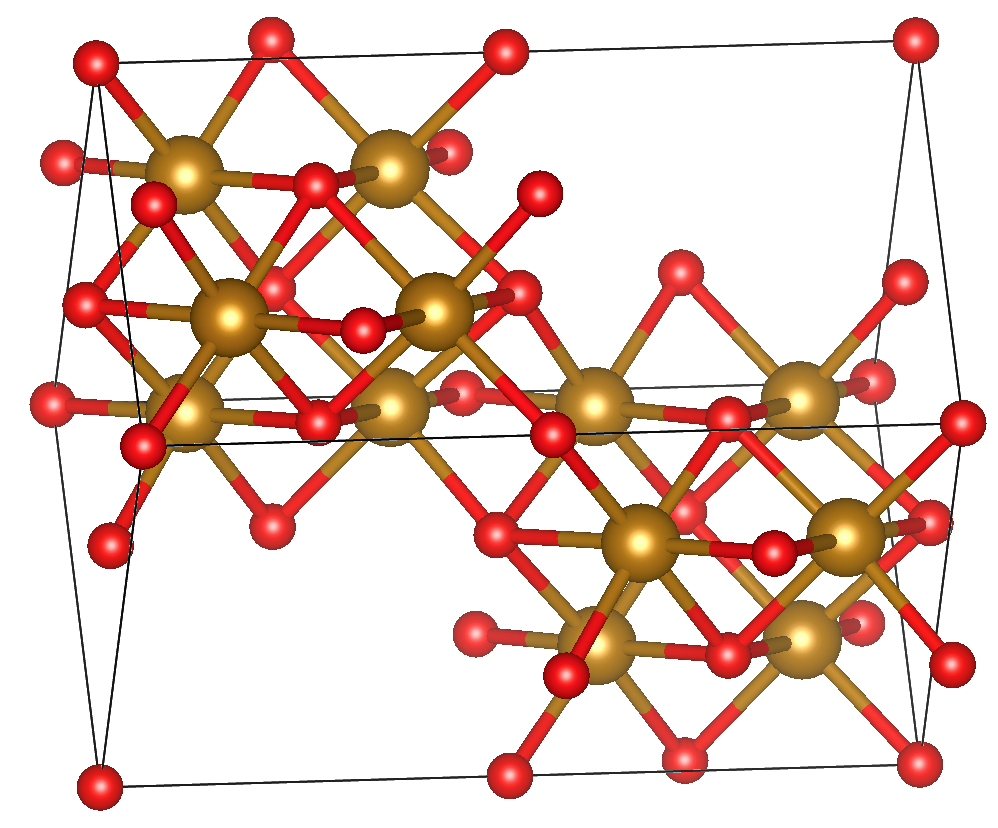}
		\caption{Graph representation of the crystal Fe\textsubscript{12}O\textsubscript{12}}
		\vspace{3pt}
	\end{subfigure}
	\caption{Cif and graph representation of crystal materials}
	\label{fig:crystal_structure}
\end{figure}


A periodic crystal structure can be represented by its lattice constants a,b,c and angles $\alpha$, $\beta$, and $\gamma$, the space group, and the coordinates at unique Wyckoff positions. Using a threshold 3.0 {\AA}, the crystal structure can be converted into a graph, which can be represented as an adjacency matrix, or contact map. To get more controlled experiments, in our experiments, we used the thresholds used by the VESTA software to define the contact map.  The contact map captures the interactions among atoms in the unit cell, which can be predicted by the known interaction patterns of these atom pairs in other known crystal materials structures. Here we assume that the perfect atom contact maps have been obtained, and we'd like to check if the global optimization algorithms can help reconstruct the crystal structures in terms of the atom coordinates from the contact map, with or without adding other geometric or physical constraints. By formulating the contact map based CSP as an optimization problem, it allows us to evaluate how different global optimization algorithms such as genetic algorithms (GA), particle swarm optimization (PSO), differential evolution (DE)  can solve this problem and how difficult this reconstruction problem is for different crystal structures of varying complexity in terms of the number of unique Wyckoff positions(which determine the number of independent variables to optimize), the level of symmetry as represented by the space group, and also the number of atoms in the unit cell, which determines the number of contact constraints. For the example in Figure\ref{fig:crystal_structure}, the number of variables to optimize is 4x3=12, corresponding to 4 Wyckoff positions each with x,y,z three coordinate values. The crystal has 24 atoms in the unit cell, which can be mapped into a 24x24 contact map matrix. The optimization problem is then how to search appropriate Wyckoff position atom coordinates so that after symmetry operations specified by space group 190, the generated crystal structure will have the same contact map matrix. In this study, we assume the space group information, and the unit cell parameters of the target composition are all known, which is reasonable as they can be predicted using different approaches \cite{song2020machine,liang2020cryspnet,jiang2006prediction,nait2020prediction}. While only contact map information is used as optimization target, other atomic interaction information such as limits of distances or preferential neighborhood relationships (e.g. atoms of some element pairs cannot stay too close to each other in known crystals) between some atom pairs can also be added as constraints in global search. The geometric constraint optimization objective can also be combined with the traditional free energy objective to achieve a synergistic effect by e.g. reducing the number of DFT free energy calculations.

\begin{figure}[ht]
  \centering
  \includegraphics[width=0.85\linewidth]{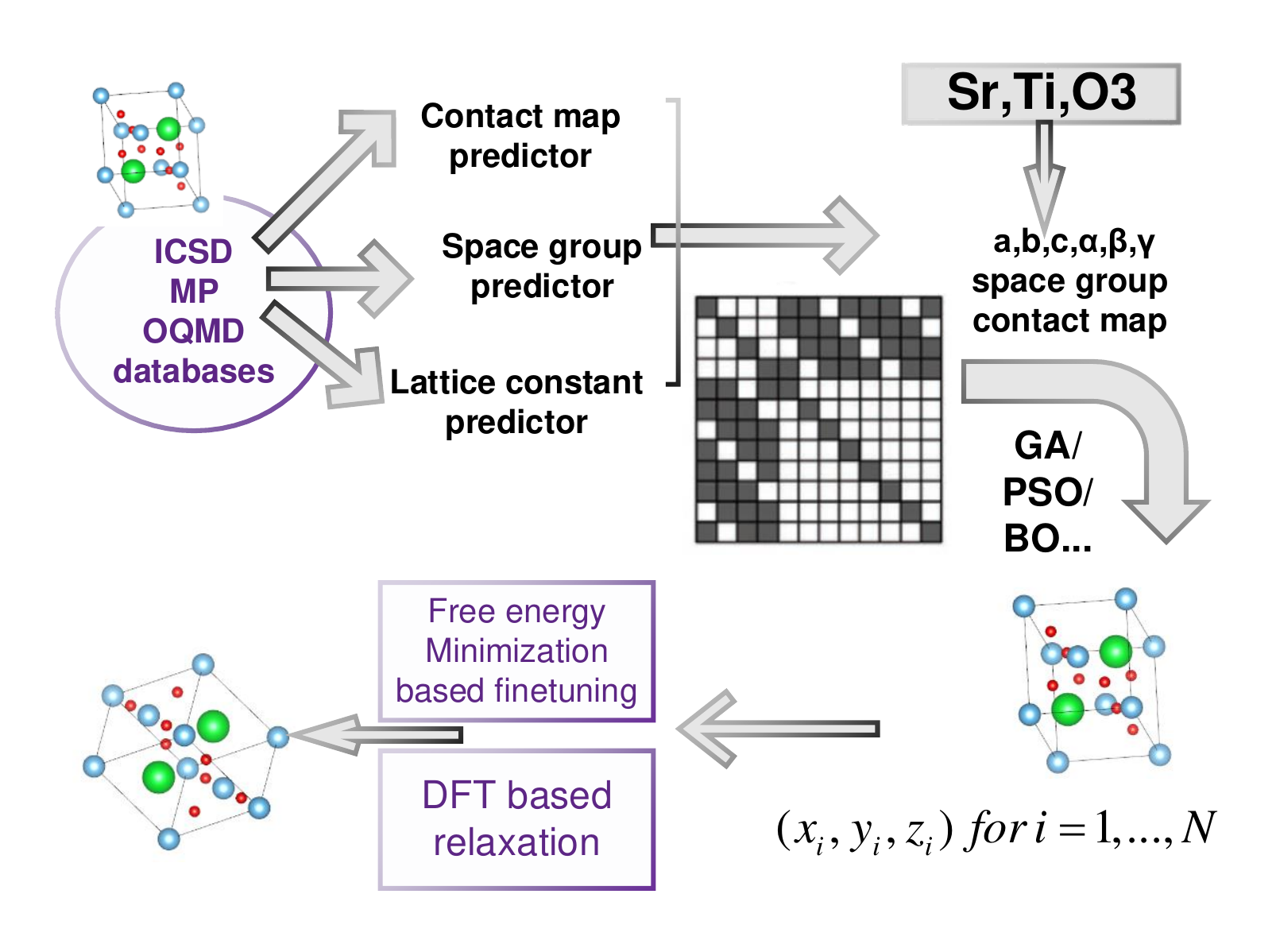}
  \caption{The CMCrystal algorithm for contact map based crystal structure prediction.}
  \label{fig:framework}
\end{figure}

\subsection{Contact map based CSP using global optimization}
In our problem formulation, the independent variables are a set of fractional coordinates $(x_i,y_i,z_i)$ for i=0,...,N, where N is the number of Wyckoff positions and $x_i,y_i,z_i$ are all real numbers in the range of [0,1]. To solve this crystal structure reconstruction problem, we propose to employ global optimization algorithms such as GAs and PSO to search the coordinates by maximizing the match between the contact map of the predicted structure and the contact map of the target crystal structure. This CMCrystal CSP framework is given in Figure\ref{fig:framework}. Basically, first, using the existing inorganic materials samples in the databases such as ICSD, Materials Project, and OQMD, three prediction models will be trained including a space group predictor \cite{zhao2020machine,liang2020cryspnet}, a lattice constant predictor\cite{zhang2020machine,nait2020prediction,jiang2006prediction,javed2007lattice,majid2010lattice}, and a contact map predictor. And then given these information a global optimization algorithm, such as the genetic algorithm, particle swarm optimization, or Bayesian optimization, to search the atom coordinates such that the resulting structure's topology(contact map) matches the predicted contact map as much as possible. After that, the structures will then be fed to free energy minimization based DFT relaxation or refinement to generate the final structure prediction. 

In this work, we focus on exploring how global optimization can be used to search the atom coordinates guided by a given contact map. We apply a set of six state-of-the-art global optimization algorithms to different problem instances to evaluate and compare their performance. Here, we summarize the main ideas of the selected optimization algorithms, their advantages, and key hyper-parameters.


\subsubsection{Genetic algorithms(GA)}

Genetic algorithms\cite{goldberg1988genetic} are population-based search algorithms inspired by the biological evolution process. Candidate solutions (individuals) are encoded by binary or real-valued vectors. Starting with a random population of individuals, the population is then subject to generations of mutation, crossover, and selection to evolve the population toward individuals with high fitness, evaluated by the optimization objective functions. Compared to other heuristic search algorithms, GAs have proved to be suitable for large-scale global optimization problems\cite{whitley2019next} and has been used in several crystal structure prediction algorithms \cite{glass2006uspex,curtis2018gator,avery2019xtalopt}, and mainly for free energy minimization. The main hyper-parameters include the population size, crossover and mutation rates. Here we apply the real-value encoded GA as the global optimization procedure for crystal structure reconstruction.

\subsubsection{Differential evolution (DE)}

Differential evolution \cite{price2006differential} is a stochastic, population-based evolutionary optimization algorithm designed for optimize real parameter, real-valued functions, many of which are nondifferentiable, non-continuous, non-linear, noisy, flat, multi-dimensional or have many local minima, constraints or stochasticity. While genetic algorithms more focus on the crossover operator, DE mainly uses its special mutation operator, which generates new candidates by adding a weighted difference between two population members to a third member. This mutation operator has an inherent adaptive characteristic to make smaller mutations when the population approaches global or local optima. It is thus usually robust and has fast convergence. It has three main parameters: the population size (usually 5-10 times of the number of variables), the scaling factor F, and the crossover rate.

\subsubsection{Particle Swarm Optimization (PSO)}
Particle swarm optimization (PSO) \cite{poli2007particle} is a population-based stochastic optimization algorithm inspired by the social behavior of some animals such as flocks of birds or schools of fish to solve nonlinear global optimization problems. The algorithm seeks the optimal solution through collaboration and information sharing between searching individuals in the group. Each individual updates its movement through the search space by combining some aspect of its own history of current and best (best-fitness) locations with those of one or more members of the swarm. Although PSO can fall into the local optimum for complexity problems, its search speed is fast, efficient, and the algorithm is simple, so 
it is used in the article to optimize the target.

\subsubsection{Bayesian Optimization (BO)}

Bayesian optimization (BO) \cite{mockus1994application,lizotte2008practical} is an algorithm for optimizing expensive objective functions that take a long time to evaluate. It is good for optimization over continuous domains of less than 20 dimensions\cite{frazier2018tutorial}. 
Bayesian optimization is one of the most efficient approaches to optimization in terms of the number of function evaluations by incorporating problem belief about the problem to help direct the sampling and by employing an automated mechanism to trade-off exploration and exploitation of the search space based on its acquisition function based sampling. Common acquisition functions include expected improvement, entropy search, and knowledge gradient. It usually uses Gaussian process regressor or deep neural networks\cite{snoek2015scalable} to build a surrogate model \cite{vu2017surrogate} for the expensive objective function and uses Bayesian estimation to calculate the prediction uncertainty at each sampling points. BO has been widely used in tuning hyper-parameters for machine learning algorithms \cite{snoek2012practical}and active learning for materials design \cite{lookman2019active}.

\subsubsection{Covariance Matrix Adaptation-Evolution Strategy (CMA-ES)}
Covariance Matrix Adaptation-Evolution Strategy (CMA-ES)\cite{hansen2006cma} is a random, fast and robust local search algorithm that does not need to calculate gradients. It samples new candidate solutions from the multivariate normal distribution of its mean, and adapts after each iteration. CMA-ES is mainly used to solve nonlinear and non-convex optimization problems. It belongs to a category of evolutionary algorithms and has randomness. Compared with most other evolutionary algorithms, it is a quasi-parameterless algorithm. CMA-ES is one of the most effective methods to deal with difficult numerical optimization problems \cite{hansen2010comparing}. CMA-ES has been widely used in practical problems \cite{bayer2004evolutionary,damp2005gonzalez,li2007comparative}. This algorithm is superior to all other similar learning algorithms in the benchmark multimodal functions. Good results with CMA-ES can be achieved when given a very large evaluation budget \cite{waibel2019building}.

\subsubsection{RBF Model-based optimization (RBFOpt)} 

RBFOpt \cite{costa2018rbfopt} is a continuous optimization algorithm based on the Radial Basis Function method. It constructs and iteratively refines a surrogate model of the unknown objective function and exploits a noisy but less expensive surrogate model to accelerate convergence to the optimum of the exact oracle. In this aspect, it shares some principles with the Bayesian optimization approach. It also introduces an automatic model selection phase during the optimization process. One of its key ideas is to use RBF interpolation to build a surrogate model, and define a measure of “bumpiness”. Given a target
objective function value at a sampling point, its bumpiness measures the likelihood that the target function value occurs there, based on the interpolation points. The assumption is that the unknown function f does not oscillate too much so that a model that can explain the data and minimizes the bumpiness can be found. Previous benchmark studies show that this algorithm has high efficiency in terms of the number of evaluations and robustness.


\subsection{Objective function and Evaluation Criteria}

The objective function for contact map based structure reconstruction is defined as the dice coefficient, which is shown in the following equation:

\begin{equation}
\operatorname{fitness}_{opt}=\operatorname{Dice}=\frac{2|A \cap B|}{|A|+|B|} \approx\frac{2 \times A \bullet B}{\operatorname{Sum}(A)+\operatorname{Sum}(B)}
\end{equation}

where$A$ is the predicted contact map matrix and $B$ is the true contact map of a given composition, both only contain 1/0 entries.  $A \cap B$  denotes the common elements of A and B, |g| represents the number of elements in a matrix, • denotes dot product, Sum(g) is the sum of all matrix elements. Dice coefficient essentially measures the overlap of two matrix samples, with values ranging from 0 to 1 with 1 indicating perfect overlap. We also call this performance measure the contact map accuracy.

To evaluate the reconstruction performance of different algorithms, we can use the dice coefficient as one evaluation criterion, which however does not indicate the final structure similarity between the predicted structure and the true target structure. To address this, we define the root mean square distance (RMSD) and mean absolute error (MAE) of two structures as below:
\begin{equation}
    \begin{aligned}
\mathrm{RMSD}(\mathbf{v}, \mathbf{w}) &=\sqrt{\frac{1}{n} \sum_{i=1}^{n}\left\|v_{i}-w_{i}\right\|^{2}} \\
&=\sqrt{\frac{1}{n} \sum_{i=1}^{n}\left(\left(v_{i x}-w_{i x}\right)^{2}+\left(v_{i y}-w_{i y}\right)^{2}+\left(v_{i z}-w_{i z}\right)^{2}\right)}
\end{aligned}
\end{equation}

\begin{equation}
        \begin{aligned}
\mathrm{MAE}(\mathbf{v}, \mathbf{w}) &=\frac{1}{n} \sum_{i=1}^{n}\left\|v_{i}-w_{i}\right\| \\
&=\frac{1}{n} \sum_{i=1}^{n}\left(\|v_{i x}-w_{i x}\|+\|v_{i y}-w_{i y}\|+\|v_{i z}-w_{i z}\|\right)
\end{aligned}
\end{equation}

where $n$ is the number of independent atoms in the target crystal structure. For symmetrized cif structures, $n$ is the number of independent atoms of the set of Wyckoff equivalent positions. For regular cif structures, it is the total number of atoms in the compared structure. $v_i$ and $w_i$ are the corresponding atoms in the predicted crystal and the target crystal structure. It should be pointed out that in the experiments of this study, the only constraints for the optimization is the contact map, it is possible that the predicted atom coordinates are oriented differently from the target atoms in terms of coordinate systems. To avoid this complexity, we compare the RMSD and MAE for all possible coordinate systems matching such as (x,y,z -->x,y,z), (x,y,z -->x,z,y), etc. and report the lowest RMSD and MAE.

\section{Experiments}
\subsection{Test problems}

We have selected a set of target crystal structures as test cases for evaluating the proposed contact map based crystal structure reconstruction algorithm using different global optimization algorithms. The list of target materials are shown in Table \ref{table:target_structures}. Here, the numbers of independent atom sites are 2 and 3 corresponding to 6 and 9 number of optimization variables. The space group numbers range from 4 to 61 corresponding to triclinic, monoclinic, and orthorhombic structurs (More symmetric structures are reported in Section 3.3.3. 

\begin{table}[H] 
\begin{center}
\caption{Statistics of target crystal structures}
\label{table:target_structures}
 \begin{tabular}{|c c c c c c|} 
  \hline
\textbf{Target} & MP\_id & \textbf{No.of  sites} & \textbf{\#Atom in unit cell} & \textbf{Space Group}& \textbf{\#variables} \\ [0.5ex] 
 \hline\hline
Ag\textsubscript{4}S\textsubscript{2}  & mp-560025  & 3  & 6  & 4 & 9  \\  \hline
Bi\textsubscript{4}Se\textsubscript{4}  & mp-1182022 & 2  & 8  & 14  & 6  \\  \hline
B\textsubscript{4}N\textsubscript{4}  & mp-569655 & 2 & 8   & 14 & 6 \\  \hline
S\textsubscript{4}N\textsubscript{4}  & mp-236 & 2 & 8  & 14 & 6  \\  \hline
Pb\textsubscript{4}O\textsubscript{4}  & mp-550714 & 2  & 8  & 29 & 6  \\  \hline
Co\textsubscript{4}As\textsubscript{8} & mp-2715   & 3  & 12   & 14   & 9   \\  \hline
Bi\textsubscript{8}Se\textsubscript{4} & mp-1102082  & 3  & 12 & 14  & 9   \\  \hline
Te\textsubscript{4}O\textsubscript{8} & mp-561224 & 3 & 12  & 19  & 9  \\  \hline
W\textsubscript{4}N\textsubscript{8}  & mp-754628 & 3  & 12  & 33  & 9  \\  \hline
Cd\textsubscript{4}P\textsubscript{8} & mp-402  & 3   & 12    & 33   & 9   \\  \hline
Ni\textsubscript{8}P\textsubscript{8} & mp-27844 & 2   & 16  & 61  & 6  \\  \hline

\end{tabular}
\end{center}
\end{table}

\subsection{Experimental Setup}



For all optimization algorithms, we set the lower boundary and upper boundary of all variables to be [0, 1] when optimizing fractional coordinates. The number of variables depends on the target materials, which is equal to the number of independent atom sites multiplied by three. For GA and DE, we set the population size to 100 and the number of generations to 1000 with a mutation probability of 0.001. For PSO, the number of particles is set as 100. For CMA-ES, we set the population size to be 300 and the generation number to be 1000. For RBFOpt, we set the max\_iterations to be 1000 and the maximum number of function evaluations in accurate mode to be 300. The population size and generation number are set based on our empirical experiments which allows us to find reasonably good structures. large population size or longer running time may further improve the results if premature convergence issue is controlled well.

\subsection{Results}

\subsubsection{Successful contact map based crystal structure predictions}

To evaluate our CMCrystal method for crystal structure prediction, we apply it to a selected set of 11 target structures as shown in Table\ref{table:target_structures} with the number of atoms ranging from 6 to 16. The total number of objective evaluations is set as 100,000. 
The overall performance of different global optimization algorithms for contact map based crystal structure reconstruction is shown in Table \ref{table:overall_performance}. We find that the contact prediction accuracy for 9 out of the 11 targets reach 100\%, demonstrating the effectiveness of our method to find the target topology from random atom coordinates using the contact map as the target.  Table\ref{table:overall_performance} also shows the RMSD and MAE of the predicted structures compared to the target structures, both of which are calculated in terms of fractional coordinates of the independent atom sites. The RMSD values range from 0.07 to 0.381 with MAE ranging from 0.054 (for B\textsubscript{4}N\textsubscript{4}) to 0.335 (for Ni\textsubscript{8}P\textsubscript{8}).

\begin{figure}[hb!]
	\centering
	\begin{subfigure}{.4\textwidth}
		\includegraphics[width=\textwidth]{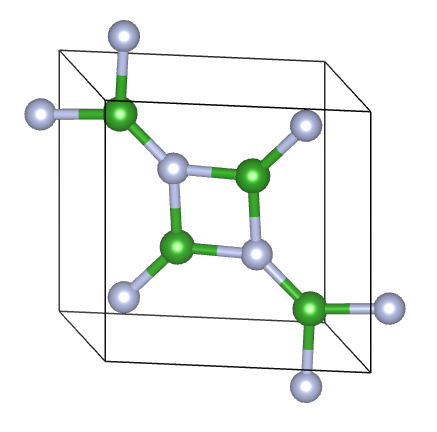}
		\caption{Target structure B\textsubscript{4}N\textsubscript{4}}
		\vspace{3pt}
	\end{subfigure}
	\begin{subfigure}{.4\textwidth}
		\includegraphics[width=\textwidth]{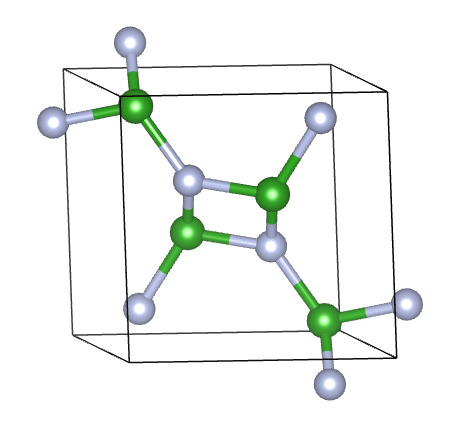}
		\caption{Predicted structure of B\textsubscript{4}N\textsubscript{4} with contact map accuracy:100\%, RMSD:0.07 }
		\vspace{3pt}
	\end{subfigure}
	\begin{subfigure}{.4\textwidth}
		\includegraphics[width=\textwidth]{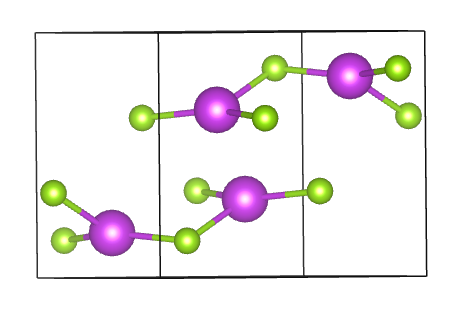}
		\caption{Target structure of Bi\textsubscript{4}Se\textsubscript{4}}
	\end{subfigure}
	\begin{subfigure}{.4\textwidth}
		\includegraphics[width=\textwidth]{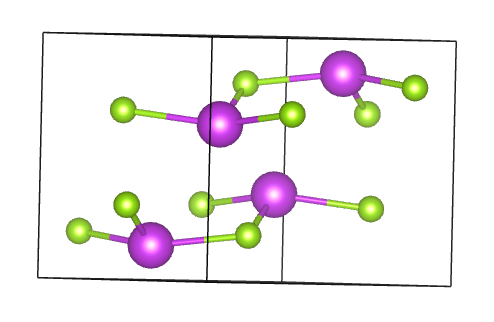}
		\caption{Predicted structure of Bi\textsubscript{4}Se\textsubscript{4} with contact map accuracy:100\%, RMSD:0.124 }
	\end{subfigure}
	
	\begin{subfigure}{.4\textwidth}
		\includegraphics[width=\textwidth]{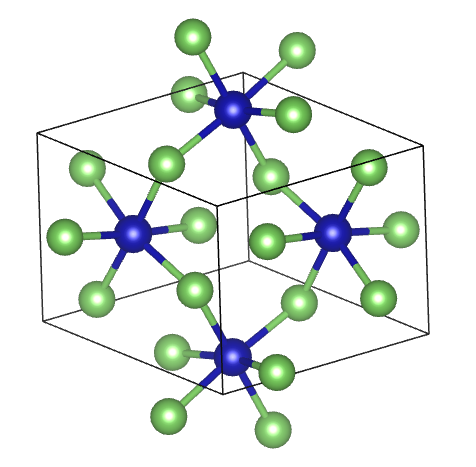}
		\caption{Target structure of Co\textsubscript{4}As\textsubscript{8}}
	\end{subfigure}
	\begin{subfigure}{.4\textwidth}
		\includegraphics[width=\textwidth]{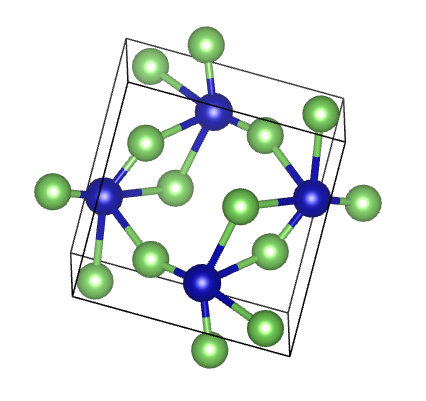}
		\caption{Predicted structure of Co\textsubscript{4}As\textsubscript{8} with with contact map accuracy:92.3\%, RMSD:0.197}
	\end{subfigure}
	
	\caption{Examples of  crystal structures predicted by CMCrystal(with GA)  versus true target structures.}
	\label{fig:predictedstructures}
\end{figure}

Figure\ref{fig:predictedstructures}  shows three sets of predicted and target crystal structures of B\textsubscript{4}N\textsubscript{4}, Bi\textsubscript{4}Se\textsubscript{4}, and Co\textsubscript{4}As\textsubscript{8}. For both B\textsubscript{4}N\textsubscript{4} and Bi\textsubscript{4}Se\textsubscript{4} (Figure\ref{fig:predictedstructures}(a)-(d)), the contact map accuracy reaches 100\% and the predicted structures are very close to the target structures. The RMSD of B\textsubscript{4}N\textsubscript{4} is 0.07 which is smaller than the RMSD (0.124) of Bi\textsubscript{4}N\textsubscript{4}, which is reflected by the higher similarity of the pairs of B\textsubscript{4}N\textsubscript{4} than the pair of structures of Bi\textsubscript{4}N\textsubscript{4}. The contact map accuracy for the target structure of Co\textsubscript{4}As\textsubscript{8} is lower with a value of 92.3\% and higher RMSD of 0.197. We note that the topology of the predicted structure in general can reach the target topology while the precise coordinates can be different.

\begin{table}[htb] 
\begin{center}
\caption{Performances of genetic algorithms in terms of contact map prediction accuracy}
\label{table:overall_performance}
\begin{tabular}{|l|l|l|l|}
\hline
\textbf{Target material} & \textbf{contact map accuracy} & \textbf{RMSD} & \textbf{MAE} \\ \hline
Ag\textsubscript{4}S\textsubscript{2}                    & 1.000            & 0.320         & 0.233        \\ \hline
Bi\textsubscript{4}Se\textsubscript{4}                   & 1.000            & 0.124         & 0.097        \\ \hline
B\textsubscript{4}N\textsubscript{4}                     & 1.000            & 0.070         & 0.054        \\ \hline
Pb\textsubscript{4}O\textsubscript{4}                    & 1.000            & 0.246         & 0.196        \\ \hline
S\textsubscript{4}N\textsubscript{4}                     & 1.000            & 0.156         & 0.137        \\ \hline
Te\textsubscript{4}O\textsubscript{8}                    & 1.000            & 0.379         & 0.266        \\ \hline
W\textsubscript{4}N\textsubscript{8}                     & 1.000            & 0.368         & 0.214        \\ \hline
Cd\textsubscript{4}P\textsubscript{8}                    & 1.000            & 0.320         & 0.204        \\ \hline
Co\textsubscript{4}As\textsubscript{8}                   & 0.923            & 0.197         & 0.149        \\ \hline
Bi\textsubscript{8}Se\textsubscript{4}                   & 0.889            & 0.257         & 0.232        \\ \hline
Ni\textsubscript{8}P\textsubscript{8}                    & 1.000            & 0.381         & 0.335        \\ \hline
\end{tabular}
\end{center}
\end{table}

\subsubsection{Performance comparison of different algorithms}

\begin{figure}[ht!]
  \centering
  \includegraphics[width=0.8\linewidth]{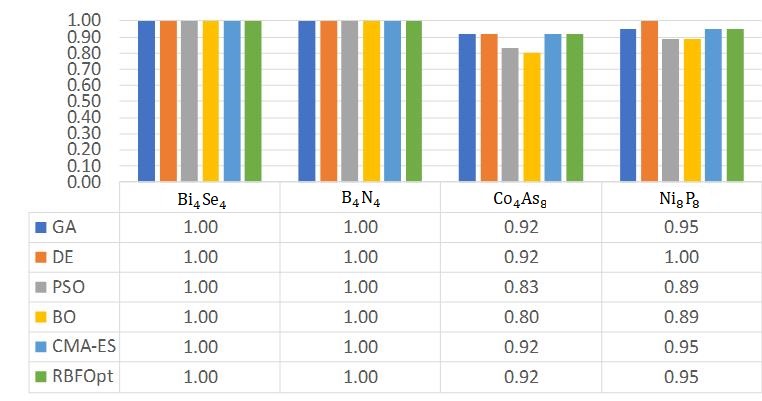}
  \caption{Performance comparison of different algorithms in terms of contact map prediction accuracy over four target structures}
  \label{fig:contactmap_accuracy}
\end{figure}

To compare the performance of different optimization algorithms for crystal structure reconstruction from contact maps, we run all six optimization algorithms for a set of target structures of different complexity. Figure\ref{fig:contactmap_accuracy} shows the performance of six algorithms. For easy cases of Bi\textsubscript{4}Se\textsubscript{4} and B\textsubscript{4}N\textsubscript{4}, all algorithms reach the 100\% accuracy for contact map prediction. For the more complex one Ni\textsubscript{8}P\textsubscript{8}, only DE achieves 100\% accuracy in the given computing budget (100,000 evaluations) while PSO and BO fall behind the most. For the most complex target Co\textsubscript{4}As\textsubscript{8}, no algorithms have achieved an accuracy of 100\% while GA, DE, CMA-ES, and RBFOpt all achieve 92\% accuracy. 

\begin{figure}[h!]
  \centering
  \includegraphics[width=0.8\linewidth]{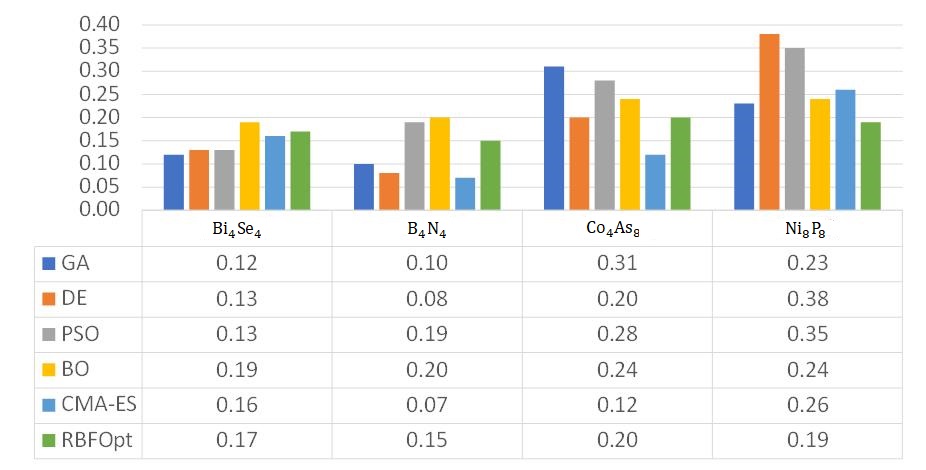}
  \caption{Performance comparison of different algorithms in terms of contact map prediction root mean square distance (RMSD) over four target structures}
  \label{fig:rsmd_performance}
\end{figure}

We also compared the RMSD performance for the six algorithms as shown in Figure\ref{fig:rsmd_performance}. Here we find that the CMA-ES achieves the best RMSD performance (0.07 and 0.12 respectively) for B\textsubscript{4}N\textsubscript{4} and Co\textsubscript{4}As\textsubscript{8}. For Bi\textsubscript{4}Se\textsubscript{4}, the best result is obtained by GA with a RMSD of 0.12. For Ni\textsubscript{8}P\textsubscript{8}, the best result is obtained by RBFOpt with a value of 0.19. However, we must note that the objective function in our study here contains only the topology information, the contact map. So algorithms with better contact map accuracy do not necessarily have better RMSD performances. In terms of computational complexity, our experiments are run on a Dell Precision workstation 
using a single CPU core with 1.7GHz. For most of the global optimization experiments here, each experiment takes about 40 minutes for results in Table 2, 70 minutes for results in Table 4 and 120 minutes for results in Table 3, which are marginal compared to the computationally demanding DFT based search algorithms.



\subsubsection{Factors that affect the optimization difficulty}

\begin{figure}[h]
	\centering
	\begin{subfigure}{.45\textwidth}
		\includegraphics[width=\textwidth]{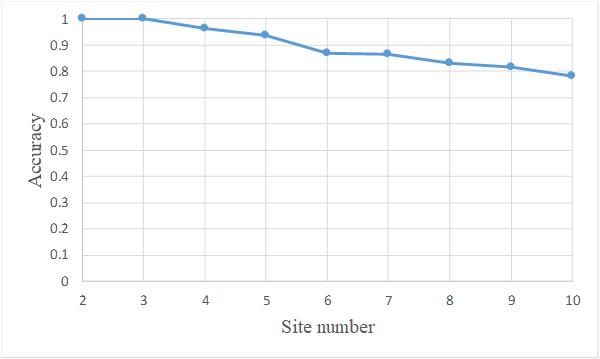}
		\caption{Contact map accuracy vs \# of atom sites}
		\vspace{3pt}
	\end{subfigure}
	\begin{subfigure}{.45\textwidth}
		\includegraphics[width=\textwidth]{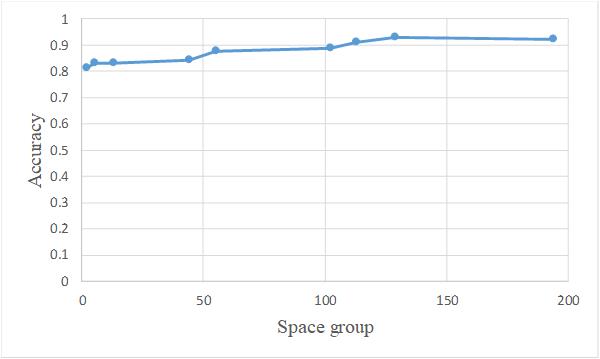} 
		\caption{Contact map accuracy vs space group}
		\vspace{3pt}
	\end{subfigure}
	
	\caption{Problem difficulty based on space group and number of atom sites}
	\label{fig:factors}
\end{figure}

From our extensive experiments, we find that there are several factors that affect the crystal structure prediction performance of our algorithms such as the number of independent atom sites, the number of atoms in the unit cell, the space group, the number of bonds/topology constraints, etc. 
Here we report how two factors, the number of independent atomic sites and the space group, affect the crystal structure reconstruction performance by the CMA-ES algorithm.  To gain a more intuitive comparison, we plot both results in Figure\ref{fig:factors}. 

In Figure\ref{fig:factors}(a), we compare the performance of CMA-ES for problem instances with the same number of total atoms in the unit cells but different numbers of independent atom sites. It shows that in general, the contact prediction accuracy gradually drops with an increasing number of atom sites, which corresponds to more optimization variables for the optimization problem. This trend is also reflected by the corresponding RMSD errors as shown in Table\ref{table:accuracy_atomsiteno}. 


Figure\ref{fig:factors}(b) and Table\ref{table:accuracy_spacegroup} show the performance results of CMA-ES structure reconstruction for a set of materials with the same number of five atom sites and similar numbers of atoms but different space groups. It is found that in general the higher the space group, the contact map accuracy is higher, indicating that higher symmetry puts more constraints on the atom configurations and reduces its search space so that better performance can be found. To be more specifically, as Table\ref{table:accuracy_spacegroup} 
shows, the contact map accuracy increases from 0.811 to 0.923 when the space group goes from 2 to 194 for Mg4Co2H10, a hexagonal structure. At the same time, the RMSD error has no consistent trend as it goes up to 0.37 and drops to 0.182 and then goes up to 0.380 and goes down to 0.276. As we discuss above, since we have only included the contact map without any distance information into our objective function, it is understandable that the RMSD have alternating up and downs. 

\begin{table}[!htb] 
\begin{center}
\caption{ Prediction performance versus number of atom sites for CMA-ES\cite{hansen2006cma}. }
\label{table:accuracy_atomsiteno}
\begin{tabular}{|l|l|l|l|l|l|}
\hline
\textbf{Target} & \textbf{mp\_id} & \textbf{atom site\#} & \textbf{atom \#} & \textbf{\begin{tabular}[c]{@{}l@{}}contact map \\ accuracy\end{tabular}} & \textbf{RMSD} \\ \hline
\hline
La\textsubscript{12}Se\textsubscript{16}    & mp-491     & 2  & 28 & 1.000 & 0.193 \\ \hline
Bi\textsubscript{8}Pd\textsubscript{4}O\textsubscript{16}   & mp-29259   & 3  & 28 & 1.000 & 0.257 \\ \hline
V\textsubscript{8}O\textsubscript{20}       & mp-25280   & 4  & 28 & 0.963 & 0.216 \\ \hline
Fe\textsubscript{12}O\textsubscript{16}     & mp-1192788 & 5  & 28 & 0.938 & 0.206 \\ \hline
Ge\textsubscript{8}S\textsubscript{12}I\textsubscript{8}    & mp-27928   & 6  & 28 & 0.870 & 0.248 \\ \hline
Li\textsubscript{4}V\textsubscript{4}Si\textsubscript{4}O\textsubscript{16} & mp-1176508 & 7  & 28 & 0.865 & 0.358 \\ \hline
Ba\textsubscript{2}V\textsubscript{8}O\textsubscript{18}    & mp-18910   & 8  & 28 & 0.831 & 0.296 \\ \hline
Si\textsubscript{2}H\textsubscript{18}C\textsubscript{6}Cl\textsubscript{2} & mp-867818  & 9  & 28 & 0.818 & 0.253 \\ \hline
Zr\textsubscript{8}Cr\textsubscript{8}F\textsubscript{12}   & mp-690241  & 10 & 28 & 0.783 & 0.349 \\ \hline
\end{tabular}
\end{center}
\end{table}

\begin{table}[htb!] 
\begin{center}
\caption{ Prediction performance versus space group using CMA-ES}
\label{table:accuracy_spacegroup}
\begin{tabular}{|l|l|l|l|l|l|}
\hline
\textbf{Target} & \textbf{mp\_id} & \textbf{atom site\#} & \textbf{space group} & \textbf{\begin{tabular}[c]{@{}l@{}}contact map \\ accuracy\end{tabular}} & \textbf{RMSD} \\ \hline
Hg\textsubscript{8}Cl\textsubscript{4}O\textsubscript{4}  & mp-636805  & 5 & 2   & 0.811 & 0.293 \\ \hline
Be\textsubscript{4}B\textsubscript{2}O\textsubscript{10}  & mp-1079124 & 5 & 5   & 0.833 & 0.371 \\ \hline
Bi\textsubscript{6}O\textsubscript{8}F\textsubscript{2}   & mp-757162  & 5 & 13  & 0.833 & 0.182 \\ \hline
Tl\textsubscript{6}V\textsubscript{2}O\textsubscript{8}   & mp-29047   & 5 & 44  & 0.842 & 0.344 \\ \hline
La\textsubscript{4}Sn\textsubscript{2}S\textsubscript{10} & mp-12170   & 5 & 55  & 0.878 & 0.335 \\ \hline
Li\textsubscript{2}V\textsubscript{2}F\textsubscript{12}  & mp-753573  & 5 & 102 & 0.889 & 0.380 \\ \hline
Yb\textsubscript{4}H\textsubscript{4}O\textsubscript{8}   & mp-625103  & 5 & 113 & 0.909 & 0.310 \\ \hline
Mg\textsubscript{4}Co\textsubscript{2}H\textsubscript{10} & mp-642660  & 5 & 129 & 0.929 & 0.376 \\ \hline
K\textsubscript{10}Cu\textsubscript{2}As\textsubscript{4} & mp-14623   & 5 & 194 & 0.923 & 0.276 \\ \hline
\end{tabular}
\end{center}
\end{table}




\section{Discussion}

Our systematic experiments have demonstrated the potential of our proposed knowledge-rich approach for crystal structure prediction. In this approach, machine learning models can be trained to predict the physical, chemical or geometric constraints such as the atomic pair contact maps and space group, and lattice parameter, which can then be used to reconstruct the coordinates of the atoms. Compared to existing ab initio free energy calculation based evolutionary search approaches, our method may achieve significant speed-up as it does not use the expensive first principle energy calculations. In many cases, the existing CSP approaches just get stuck and fail to find the stable structures. This shows the advantages of our algorithm to exploit the large number of known structures in crystal materials databases. One of the potential challenges of our algorithm is that its performance depends on the prediction accuracy of those constraints such as the contact map and lattice parameters, which may be biased by known structures. However, this issue can be addressed to certain extent by the DFT based first principle calculation of the formation energy or even thermodynamic stability phonon calculation to verify if the predicted structure is stable or not. Heuristic rules of crystal structures such as Pauling's rules can also be used to correct some errors in contact map prediction. On the other hand, the ab initio approaches have the advantages of unbiased free-energy guided global search in the configuration space, which is however usually subject to getting stuck in local optima and failure to find the true stable structures, especially for relatively large systems. Considering that the currently known ~200,000 crystal structures in the ICSD database can be classified into about 9000 structure prototypes, it means that there is a large similarity among the crystal structures that can be exploited to do CSP. Actually, the advantages and disadvantages of our knowledge-rich approach CMCrystal compared to the ab initio approaches \cite{lee2017ab} are analogous to what has been discussed in the context of the protein structure prediction field, where the prior approaches have the dominating role\cite{kuhlman2019advances}. While there is no coevolutionary information can be exploited in CSP, the physical and chemical rules that govern atom interactions allow predicting the contact with good performance.

Another factor that should be considered for CMCrystal algorithm is how well the required constraint information can be predicted. In a recent work on space group prediction\cite{liang2020cryspnet}, the top three space group prediction performance has reached a range of 0.81 to 0.98 in terms of $R^2$ scores based on different crystal systems. For cubic structures, this performance is 0.96 on average. For the lattice parameter prediction, it has achieved a performance of around 0.82 while our in-house algorithm has reached a $R^2$ of 0.97 for cubic structures and a $R^2$ range of 0.77 to 0.89 for other five crystal systems. Based on current performance and ongoing efforts in this area, we believe that our proposed algorithm can achieve good performance for quite some types of crystal materials. 

While CMCrystal has been shown to be able to reconstruct atomic configurations from the contact map in this study, the quality depends on how accurate is the predicted contact map, which is an unsolved problem to be studied. However, considering the strong consistency of the bond lengths among atom pairs of different species in different compounds, the contact maps of crystal materials are expected be predictable. Actually, one of our ongoing works have focused on the development of deep neural network models for contact map prediction and has achieved very promising results.

\FloatBarrier
\section{Conclusion}

We formulate a crystal structure prediction/reconstruction problem based on its space group symmetry and the atom contact map, and applied a series of state-of-the-art global optimization algorithms to solve the problem. Our experiments show that global optimization algorithms can reconstruct the crystal structure for some materials by optimizing the placement of the atoms using the contact map as the objective given only their space group and stoichiometry. These predicted structures are close to the target crystal structures so that they can be used to seed the costly free energy minimization based crystal structure prediction algorithms for further structure refining. They may also be used for DFT based structure relaxation to obtain the correct crystal structures for some compositions. However, we found that using the contact map alone is in general not enough to guide the search for the true structure precisely and additional geometric and physical constraints may be needed such as pairwise distance information to further improve the reconstruction quality, which is under our investigation. Another potential improvement is to conduct more extensive parameter tuning for the optimization algorithms used here for different structures as here we mostly use the default parameters for the algorithms. 


\section{Availability of data}

The data that support the findings of this study are openly available in Materials Project database at http:\\www.materialsproject.org 

\section{Contribution}
Conceptualization, J.H.; methodology, J.H. and W.Y.; software, W.Y. and J.H; validation, W.Y, J.H., E.S., R.D., Y.L.;  investigation, J.H., W.Y., R.D., E.S., and S.L.; resources, J.H.; data curation, J.H. and W.Y.; writing--original draft preparation, J.H., R.D., Y.L, W.Y. and X.L.; writing--review and editing, J.H and R.D.; visualization, J.H., R.D., and W.Y; supervision, J.H.;  funding acquisition, J.H.

\section{Acknowledgement}
Research reported in this work was supported in part by NSF under grant and 1940099 and 1905775. The views, perspective, and content do not necessarily represent the official views of NSF.

\bibliography{references}
\bibliographystyle{unsrt}

\end{document}